\documentclass[conference]{IEEEtran}

\usepackage{float}
\usepackage{subfig}
\usepackage{amsmath,epsfig,psfrag} %subfigure
\usepackage{amsfonts,color,graphics,graphicx}
\usepackage{algorithm} %,algorithmic
\usepackage{algpseudocode}
\usepackage{multirow}
\usepackage{rotating}
\usepackage[numbers]{natbib}
\usepackage[numbers]{natbib}
\usepackage[normalem]{ulem}
\renewcommand{\baselinestretch}{1}

\usepackage{verbatim}

\begin{document}

\title{Is Uncertainty Always Bad?: Effect of Topic Competence on Uncertain Opinions}

\author{\IEEEauthorblockN{Jin-Hee Cho\IEEEauthorrefmark{1} and
Sibel Adal{\i}\IEEEauthorrefmark{2}}
\IEEEauthorblockA{\IEEEauthorrefmark{1}US Army Research Laboratory, Adelphi, MD\\ Email: jin-hee.cho.civ@mail.mil}
\IEEEauthorblockA{\IEEEauthorrefmark{2}Rensselaer Polytechnic Institute, Troy, NY \\Email: sibel@cs.rpi.edu}}

% make the title area
\maketitle

\begin{abstract}
The proliferation of information disseminated by public/social
media has made decision-making highly challenging due to
  the wide availability of noisy, uncertain, or unverified
information. Although the issue of uncertainty in information
has been studied for several decades, little work has investigated how
noisy (or uncertain) or valuable (or credible) information can be
formulated into people's opinions, modeling uncertainty both in the
quantity and quality of evidence leading to a specific opinion.
In this work, we model and analyze an opinion and information
model by using Subjective Logic where the initial set of evidence is
mixed with different types of evidence (i.e., pro vs. con or noisy
vs. valuable) which is incorporated into the opinions of original
propagators, who propagate information over a network. With the
help of an extensive simulation study, we examine how the different
ratios of information types or agents' prior belief or topic competence affect the overall information diffusion. Based on our findings, agents' high uncertainty is not
necessarily always bad in making a right decision as long as they are
competent enough not to be at least biased towards false information
(e.g., neutral between two extremes).
\end{abstract}

\vspace{-2mm}
\section{Introduction}
\vspace{-2mm} 
Decision making under uncertainty becomes more
challenging as we are living with flooding amounts of unverified
information provided by public/social media,
and/or interpersonal interactions in an individual's online/offline
social network. Many existing studies have examined
the key factors that impact the dynamics of opinions in different
domains. In the mass communication, perceptions of news
credibility from public media (e.g., newspapers, online news,
television, radio) have been studied to investigate if there exists
any link between the change of public opinion and the perception of
public media news~\cite{Kiousis01}. In the network science, a rich
amount of literature investigated how the choice of seeding nodes
propagating initial information affects information
diffusion~\cite{Cho14, Li15}. However, little work has studied how
individuals process uncertain, noisy information, how their topic
competence (e.g., expertise or knowledge on a given topic) affects
processing of noisy (or uncertain) or valuable (or credible)
information, and the subsequent results of these factors on the
opinions and beliefs of individuals.

In this work, we develop an agent-based opinion and information model
by considering agents processing information differing in its bias
  (e.g. supporting a pro or con point of view) and in its value
  (e.g. noisy vs. valuable) and forming opinions with a degree of
  uncertainty.
In this work, we define the concepts of `noisy' and `valuable' in
terms of uncertainty and credibility, respectively. In particular, the
concept of uncertainty indicates {\em vagueness} (i.e., unclear
without reasonable, analytical reasoning or clear facts) and/or {\em
  ambiguity} (i.e., inconsistency with conflicting evidence) which may
not be useful for agents to collect credible evidence supporting
either pro or con. On the other hand, the concept of credibility
refers to conciseness and consistency that can provide a clear stance
by providing clear, factual details based on a rational, analytical
reasoning process~\cite{Hughes15}. However, credible information is not
necessarily true as we can find information that looks credible and
real but it turns out to be false. This is why we consider the four
types of information, given either pro or con can be true. In
  cases where there is a clear true point of view, we declare one of
  pro vs. con points of view to be correct.  While there exists work in belief models (e.g., Subjective Logic) to tie uncertainty in beliefs to both quantity and balance of information supporting pro or con points of view, there is no model that also incorporates quality of information and an agent's ability to assess quality into opinion models. However, the interplay between these two dimensions is a crucial aspect of dissemination of biased and incorrect information in today's social networks and has not been widely studied.
  
%  This work does not deal with text mining to identify information credibility or uncertainty, but rather focuses on modeling and analysis of the proposed opinion and information model as detailed in Section \ref{sec:opinion_model}.

This work has the following {\bf key contributions}:
\begin{itemize}
\vspace{-1mm}
\item We develop an agent-based opinion and information model which
  allows the agent to form a uncertain opinion based on Subjective
  Logic (SL) and to process information different along two
    different dimensions in order to explicitly deal with uncertainty in its opinion. Our model
  allows four types of information, modeled along with {\em pro (P)
    vs. con (C)} and {\em noisy (N) vs. valuable (V)} dimensions, as
  detailed in Section \ref{subsec:information_model}. To the best of
  our knowledge, no prior work has considered the formulation of
  SL-based opinions that can accept these types of information which
  can be easily found in reality.
\item We model agents' topic competence which can significantly affect
  their information processing in diagnosing given evidence on whether
  it is noisy or valuable and their decisions on whether to accept
  false information after interacting with other agents. This allows
  us to analyze the effect of agents' topic competence on the degree
  of agents' belief in false information under the various
information scenarios.
\item We conduct comprehensive simulation experiments in order to
  identify key factors that can increase the fraction of agents not
  believing in false information, given that `belief' in SL represents false information. Our findings show that
  uncertainty is not necessarily always bad in making a right decision
  when agents are capable of not being biased for false information
  such as having a neutral stance in their prior belief/disbelief towards
    the given topic.
\end{itemize}

\vspace{-3mm}
\section{Background \& Related Work} \label{sec:related_work}
\vspace{-1mm} 
In this section, we discuss existing work on
decision making under uncertainty and information credibility in
public/social media. We also give a brief overview on approaches
to mitigate or remove false information propagation in social
networks.
\vspace{-3mm}
\subsection{Uncertainty in Decision Making} \label{subsec:related_uncertainty_dm}
\vspace{-1mm} 
Many state-of-the-art studies have investigated how decision makers
use other alternative methods to deal with uncertainty
in decision making such as herding~\cite{Avery98, Fischer86, Rosas17} or cognitive bias~\cite{Norman12, Tversky74}.
However, no prior work has investigated how an individual agent forms
its uncertain opinion in processing information composed of credible
(or valuable) or uncertain (or noisy) evidence, which is explored in
this work.

\vspace{-3mm}
\subsection{Information Credibility in Public/Social Media}
\vspace{-1mm} Many approaches to derive information credibility have
been studied using social media datasets. \citet{Castillo11} developed
a credibility methodology by identifying a set of features
representing false information propagated through Twitter.
\citet{Sikdar13} proposed a methodology that combines both
  direct and heuristic evaluations of credibility with superior
  performance to the baseline counterparts. However, these
  studies do not model or track change in opinions of individuals and
  their uncertainty given a ground truth of information credibility.
%use the human assessed
%results as the ground truth of information credibility in which
%humans' imperfect judgment errors have not paid attention.
\vspace{-3mm}
\subsection{Countering False Information Propagation}
\vspace{-1mm} Existing approaches to counter false information use two
main approaches: \emph{network analysis-based approach} (NA) and
\emph{feature-based approach} (FA). NA aims to stop false information propagation by selecting a set of counter-misinformation nodes (i.e., nodes to disseminate true information against misinformation)~\cite{Nguyen12}. The process of misinformation is also modeled based on epidemic models with variants of SIR (Susceptible-Infected-Recovered) model by introducing `forgetting/remembering factor' in rumor propagation~\cite{Zhao11}. FA concentrates on identifing key features of false information or the sources (i.e., users) of the false information in gossip diffusion~\cite{Wang15} and feature extraction for rumor identification~\cite{Kumar14}.

Unlike the existing approaches in both NA and FA, this work examines
uncertainty of opinions in analyzing the propagation of information
considering agents' topic competence, prior belief and their opinion
updates based on information composed of the four different types as
discussed earlier. 
\vspace{-2mm}
\vspace{-3mm}
\section{Opinion Model} \label{sec:opinion_model}
\vspace{-1mm}
In SL, an \emph{opinion} is represented by three dimensions: \emph{belief} ($b$), \emph{disbelief} ($d$), and \emph{uncertainty} ($u$)~\cite{Josang16}. A single opinion on a given proposition is represented by:
\vspace{-1mm}
\begin{eqnarray}
{b, d, u} \in [0, 1]^3, & b + d + u = 1
\vspace{-3mm}
\end{eqnarray}
We adopt SL to explicitly deal with uncertainty an agent perceives in updating its opinion.
\vspace{-2mm}
\subsection{Opinion Formation}
\vspace{-1mm}
The agent may have degrees of belief (i.e., agree or pro) and/or disbelief (i.e., disagree or con) towards a given proposition with some degree of uncertainty. Agent $i$'s opinion on proposition $A$ is denoted as $w_i^A$. For simplicity, we omit $A$ and use $w_i$ to represent an agent $i$'s opinion as $w_i = \{ b_i, d_i, u_i, a_i \}$ where $a_i$ is the base rate which normally represents general background knowledge or judgment bias~\cite{Josang16}. %In this work, we interpret $a_i$ reflecting the prior belief which is generally accepted by society. For example, if the society is more conservative, we assume the general public is more favored for conservative decisions, and vice versa.

The base rate, $a_i$, affects expectation probability (i.e., a probability that an agent is expected to make a decision) in either belief or disbelief~\cite{Josang16}, denoted by $E_{b_i}$ or $E_{d_i}$, respectively, where they are given by:
\vspace{-2mm}
\begin{eqnarray} \label{eq:expectation}
\vspace{-2mm}
\small
E_{b_i} = b_i + a_i u_i, \;
E_{d_i} = d_i + (1-a_i) u_i.
\vspace{-2mm}
\end{eqnarray}
Note that $E_{b_i}+E_{d_i}=1$ as $b_i + d_i + u_i=1$. An individual's acceptance towards given information is
affected by various factors, including personality (e.g.,
agreeableness, open-mindedness, stubbornness), impact of neighbors
(e.g., herding), homophily (e.g., like-mindedness),
competence (e.g., domain knowledge), or confidence (e.g., certain
about its own opinion). In this work, we particularly model and
analyze the effect of individuals' topic competence on a given
proposition in which the topic competence may adjust the degree of the
prior belief and affect final decisions.

In SL, an agent forms its opinion based on the amount of directly observed evidence based on the following mapping rule:
\vspace{-2mm}
\begin{eqnarray} \label{eq: mapping}
\vspace{-2mm}
\small
b = \frac{r}{r+s+W}~, 
d = \frac{s}{r+s+W}~, 
u = \frac{W}{r+s+W}~.
\end{eqnarray}
where $r$ is positive evidence and $s$ is negative evidence for a particular proposition. For simplicity, we dropped the subscript $i$ denoting the agent. When $W=0$, $b$ is a natural estimate of the fractional evidence in favor of the proposition. To be specific, $W$ indicates the amount of uncertainty introduced by the inherent errors or imperfect observability. 
\vspace{-3mm}
\subsection{Opinion Update} \label{subsec:opinion_update}
\vspace{-2mm}
In this section, we describe how our work modeled the following: (i) the impact of an agent's topic competence on its prior belief/disbelief; (ii) homophily-based (i.e., like-minded) opinion update; and and (iii) opinion decay over time. We adopt the features of (ii) and (iii) from our prior work~\cite{Cho17} and include them here to be self-contained.
\subsubsection{Prior Belief based on Topic Competence}
An agent's topic competence is defined as the critical thinking
  ability which can be against bias or prior belief even if
  such belief is generally accepted by others in the
  network~\cite{Stanovich97}. As this work models this characteristic
as a parameter, each agent $i$ is characterized by its own topic
competence $tc_i$, ranged in $[0, 1]$ as a real number, which is used
for agent $i$ to judge if given information is valuable or noisy or if the prior belief or prior disbelief, $a_i$ or $1-a_i$, is properly supporting given information (e.g., given information is true or false). $tc_i$ also affects how much agent $i$ is for/against its prior belief/disbelief favoring pro or con, $a_i$ or $1-a_i$, depending on its judgmental ability to diagnose true/false information. Assuming that given belief, $b_i$, represents the degree that $i$ believes false information, $i$'s prior belief can be adjusted based on its competence $tc_i$ by:
\begin{equation} \label{eq:base}
\vspace{-1mm}
\small
\hat{a}_i = (1-(tc_i-0.5)) a_i
\vspace{-1mm}
\end{equation} 
where $\hat{a}_i$ is the adjusted prior belief while the prior disbelief is $(1-\hat{a}_i)$. Note that this is the case assuming that a given belief represents the support for false information which is denoted by $pro$ while a given disbelief indicates not believing the false information, denoted by $con$ in the scenario considered in this work. Recall that the adjusted prior belief/disbelief, $\hat{a}_i$/$(1-\hat{a}_i)$, will replace $a_i$/$(1-a_i)$ in Eq. \eqref{eq:expectation}, which significantly impacts agents' final decision in pro/con. Eq. \eqref{eq:base} implies that when agent $i$ has high competence, $i$ is
less likely to believe in false information. Note that $tc_i=0.5$ implies that the original prior belief is used as intact (i.e., $tc_i = 0.5$) and there is a chance for the prior belief to increase when $tc_i < 0.5$. This indicates that when an agent has below the average competence (i.e., $tc_i = 0.5$), an agent's bias can be even more pronounced. On the other hand, its bias can be relaxed with the high competence when it is above the average competence (i.e., $tc_i > 0.5$).
\subsubsection{Homophily-based Opinion Consensus}
Homophily, or like-mindedness, significantly affects the way people
update opinions~\cite{Li11}. In this work, the similarity of two
agents' opinions, denoted by $s_i^j$, is computed based on
\emph{cosine similarity} of the two opinion vectors $i$
and $j$ in terms of their belief and disbelief, $(b_i, d_i)$ and
$(b_j, d_j)$, respectively. The calculation of $s_i^j$ based on the
cosine similarity is omitted due to its popularity and space constraint.

We use $s_i^j$, as a discounting operator~\cite{Josang16}, to
determine the degree to which agent $i$ accepts agent $j$'s
opinion. Given two vectors of opinions, $w_i=\{b_i, d_i, u_i\}$ and
$w_j=\{b_j, d_j, u_j\}$, agent $j$'s trust opinion in $i$'s opinion,
$s_i^j$, is given by $w_{i \otimes j} = \{b_{i \otimes j}, d_{i
  \otimes j}, u_{i \otimes j}\}$ where each element is estimated by:
\begin{eqnarray} \label{discounting}
\vspace{-3mm}
\small
b_{i \otimes j}  =  s_i^j b_j,~  
d_{i \otimes j}  =  s_i^j d_j,~ 
u_{i \otimes j}  =  1-s_i^j(1-u_j)~.
\vspace{-3mm}
\end{eqnarray}
where $u_{i \otimes j}$ is simply derived by $u_{i \otimes j}=1-b_{i \otimes j}-d_{i \otimes j}$ and $b_j+d_j+u_j=1$. For simplicity, we omit the time step notation, but both sides of the equation refer to time step $t$.

We use SL's \emph{consensus} operator~\cite{Josang16} for an agent's opinion update upon receiving new information. The updated opinion of agent $i$ after interaction with agent $j$ is denoted as $w_i \oplus b_{i \otimes j} = \{b_i \oplus b_{i \otimes j}, d_i \oplus b_{i \otimes j}, u_i \oplus b_{i \otimes j}\}$ and each element is given by:
\begin{eqnarray} \label{consensus_op}
\vspace{-3mm}
\small
b_i \oplus b_{i \otimes j} & = &  \frac{b_i  u_{i \otimes j} + b_{i \otimes j} u_i}{\beta}~, \\ \nonumber
d_i \oplus d_{i \otimes j} & = &  \frac{d_i u_{i \otimes j} + d_{i \otimes j} u_i}{\beta}~, \\ \nonumber
u_i \oplus u_{i \otimes j} & = &  1-(b_i \oplus b_{i \otimes j} + d_i \oplus d_{i \otimes j})~. 
\vspace{-3mm}
\end{eqnarray}
where $\beta = u_i + u_{i \otimes j} - u_i u_{i \otimes j}$ and $\beta \neq 0$ is assumed. We omit the time step notation; the left side represents $w_i (t+1)$ while the right side uses the opinions at $t$ such as $w_i (t)$ and $w_j (t) = w_{i \otimes j} (t)$.
\subsubsection{Opinion Decay over Time}
Unless an agent receives new information by interacting with other agents, its opinion decays over time based on a decay factor, $\gamma$, over belief and disbelief while uncertainty increases in proportion to $\gamma$. For example, human cognition is limited by forgetting information over time. We model the decayed opinion by:
\begin{eqnarray} \label{eq: decay}
\vspace{-3mm}
b_i   =  (1-\gamma) b_i,~ 
d_i   =  (1-\gamma) d_i~, 
u_i   =  u_i+\gamma(1-u_i)~.
\vspace{-3mm}
\end{eqnarray}
Note that $u_i$ is simply derived based on $1-b_i-d_i$ where
$b_i+d_i+u_i=1$. Different from the opinion update by
Eq. \eqref{consensus_op} which allows the opinion update only for
$\beta > 0$ and $u_i> 0$, respectively, the opinion decay based on
Eq. \eqref{eq: decay} occurs at every time step. Therefore, even if
$u_i$ reaches 0, over time it can increase (i.e., $u_i>0$) and
accordingly agent $i$ can update its opinion upon receiving new
information from its neighbors. For simplicity, we omitted the time
step notation, but the left side is at time $t+1$ while the right side
is at $t$. 
\subsection{Information Model} \label{subsec:information_model}
When an agent receives information from public/social media (e.g., news articles shared), it discerns if the received information is credible or not. To model Quality-of-Information (QoI) representing information credibility in the received information, we adopt the information model used in \cite{Hughes15} where QoI is presented by the following four aspects of information: (1) {\em noisy} information without any reasonable, analytical reasoning or clear facts; (2) {\em valuable} information with clear, factual details based on a rational, analytical reasoning process; (3) information to support \emph{pro} for a given proposition (e.g., believing a false rumor); and (4) information to support \emph{con} against a given proposition (e.g., disbelieving a false rumor).

Recall that valuable (or credible) information is not necessarily true. Regardless of the actual truthfulness of information, information can be noisy or valuable. We consider the four categories of information~\cite{Hughes15} as:
\begin{itemize} 
\vspace{-1mm}
\item \emph{Pro / Valuable} ($PV$): Information supports pro for given information and is valuable.
\item \emph{Pro / Noisy} ($PN$): Information supports pro for given information but is noisy.
\item \emph{Con / Valuable} ($CV$): Information supports con for given information and is valuable.
\item \emph{Con / Noisy} ($CN$): Information supports con for given information but is noisy.
\vspace{-1mm}
\end{itemize}
In this paper, we propose a novel method to incorporate this
  information model to SL.  We formulate
the QoI based on Beta distribution~\cite{Josang16} considering the
degree of uncertainty as an opinion (see Eq. \eqref{eq: mapping}),
$\mathcal{Q}=\{q_b, q_d, q_u\}$, in SL treating the amount of CN and
PN as the amount of uncertain evidence (i.e., $W$ in Eq. \eqref{eq:
  mapping}) where the number of PV as the amount of evidence to
support a pro (i.e., $r$ in Eq. \eqref{eq: mapping}), and the number
of CV as the amount of evidence to support a con (i.e., $s$ in
Eq. \eqref{eq: mapping}). In reality, a human with limited cognitive
capability cannot capture the ground truth information. How to
perceive information (e.g., noisy vs. valuable) is
significantly affected by an agent's competence, prior belief or
bias. We define an agent's perceived opinion
$\mathcal{\hat{Q}}_i=\{q_{b_i}, q_{d_i}, q_{u_i}\}$ based on its
\emph{topic competence}, $tc_i$. Note that agents' topic competence
$tc_i$ also affect the degree of their prior belief as described in
Eq. \eqref{eq:base}. The agent's perceived opinion,
$\mathcal{\hat{Q}}_i$, is estimated based on the mapping rule (Eq. \eqref{eq: mapping}) where
the number of {\em perceived} positive, negative, and uncertain
evidence by agents is obtained by Algorithm \ref{algo:
  mapping_evidence_to_opinion} which returns $n_b$, $n_d$, and $n_u$,
respectively. Given $n_s=n_b+n_d+n_u$, each element of an opinion
$\mathcal{\hat{Q}}_i$ is given by:
\vspace{-3mm}
\begin{eqnarray} \label{eq:per_qoi}
\small
q_{b_i} = \frac{n_b}{n_s}~, \; \;
q_{d_i} = \frac{n_d}{n_s}~, \; \;
q_{u_i} = \frac{n_u}{n_s}~.
\end{eqnarray}
In Algorithm \ref{algo: mapping_evidence_to_opinion}, we assume that an agent can perfectly judge whether evidence is $pro$ or $con$, but cannot perfectly know whether it is valuable or noisy. Hence, it uses its competence, $tc_i$, to judge the evidence's value (noisy or valuable).
\vspace{-1mm}
This opinion, $\mathcal{\hat{Q}}_i$, is used to initialize the opinions of original propagators who first receive public/social media information and disseminate it to its neighboring agents in a network.
\vspace{-1mm}
\begin{algorithm}
\small
\caption{Mapping Evidence to an Opinion}
\label{algo: mapping_evidence_to_opinion}
\begin{algorithmic}[1]
\Procedure{[$n_b$, $n_d$, $n_u$] = MapEvidence}{$tc$, $\mathbf{EVD}$}
\State $tc$: an agent's topic competence
\State $\mathbf{EVD}$: a matrix of evidence (4 columns for PV, PN, CV, CN and $t$ rows for time steps from $t=1, \cdots, T$) where evidence is mapped to one of the four types of evidence, it is 1; 0 otherwise (e.g., $[1, 0, 0, 0]$ represents the evidence is PV)
\State $n_b=0$: counter for \# of evidence supporting belief
\State $n_d=0$: counter for \# of of evidence supporting disbelief
\State $n_u=0$: counter for \# of of uncertain evidence, supporting neither belief nor disbelief
\State $T$: Number of evidence
\For{$t=1$ to $T$}
\State $r=rand()$ 
\Comment{a random real number in $[0, 1$] based on uniform distribution}
\If{$\mathbf{EVD}(t)==1$}
\If{$r \leq tc$} \Comment{knows evidence is valuable} 
\State  $n_b \leftarrow n_b + 1$ 
\Else 
\State $n_u \leftarrow n_u + 1$
\EndIf
\ElsIf{$\mathbf{EVD}(t)==2$}
\If{$r \leq tc$} \Comment{knows evidence is noisy}
\State  $n_u \leftarrow n_u + 1$ 
\Else 
\State $n_b \leftarrow n_b + 1$
  \EndIf
\ElsIf{$\mathbf{EVD}(t)==3$} 
\If{$r \leq tc$} \Comment{knows evidence is valuable}
\State $n_d \leftarrow n_d + 1$ 
\Else 
\State $n_u \leftarrow n_u + 1$ 
\EndIf  
\Else $\; \; \; $ \Comment{$\mathbf{EVD}(t)==4$} 
\If{$r \leq tc$} \Comment{knows evidence is noisy}
\State $n_u \leftarrow n_u + 1$ 
\Else   
\State $n_d \leftarrow n_d + 1$
\EndIf  
\EndIf  
\EndFor
\EndProcedure
\end{algorithmic}
\end{algorithm}
\section{Agent Model} \label{sec: agent_model}
\vspace{-1mm} This work considers an online social network as a
directed or undirected graph $\mathcal{G}$ where vertices, $v_i$'s,
are agents $i$'s (e.g., users) in the set of $\mathcal{V}$ and the
edges, $e_{ij}$'s (i.e., 1 for an edge and 0 for no edge), represent
the relationships in the set $\mathcal{E}$. Agent $i$'s neighbors
refer to other agents directly connected to $i$. In this work, a
network is initialized with a set of agents receiving
information from public/social media. Then, the agents propagate their
opinions based on the received information to its neighboring agents
(i.e., directly connected agents). The neighboring agents propagate
their opinions further over a network. Hence, after the initial set of
seeding agents propagate their own opinions formed based on
their competence which processes the four types of evidence,
the information propagation continues until every agent has a chance
to propagate its own opinion (i.e., not the original opinion from the
original propagator) to its neighbors. This way of the opinion
propagation is to reflect the reality that people tend to talk about
their own opinions, not necessarily what they exactly heard from
others or media upon interactions with others.
\subsection{Agent Types}
We have two types of agents as follows:
\begin{itemize}
\item {\em Originators} (Os): This agent $i$ is selected to receive media information which consists of both noisy and valuable information. Its opinion vector $\{b_i, d_i, u_i\}$ is set to $\{b_i, d_i, u_i\}=\{q_{b_i}, q_{d_i}, q_{u_i}\}$ based on Eq. \eqref{eq:per_qoi}. A set of original propagators (Os), $s^*$, is initially selected before the interactions between agents. This type of agents does not change for the whole session after they form their opinions based on Eq. \eqref{eq:per_qoi}.
\item {\em Propagators} (Ps): This agent $i$ has low confidence (i.e., $u \rightarrow 1$) in its own opinion by not initially agreeing or disagreeing with given information (i.e., $b \rightarrow 0$ and $d \rightarrow 0$). This agent is initialized with its opinion with $(r, s, W) = (1, 1, n)$ where $n >> 1$, leading to $\{b_i, d_i, u_i\} = \{\frac{1}{n+2}, \frac{1}{n+2}, \frac{n}{n+2}\}$, implying low confidence in a given proposition due to lack of information (i.e., ignorance). This type of agents keeps updating their opinions unless its uncertainty $u$ reaches 0, based on SL's consensus operator in Eq. \eqref{consensus_op}.
\end{itemize}
As discussed in Section \ref{subsec:opinion_update}, uncertainty can increase as an opinion decays over time as shown in Eq. \eqref{eq: decay}. For simplicity, we adopt the high effectiveness (influence) property~\cite{Budak11} assuming that when $j$ propagates its opinion to its neighbor $i$, then $i$ will accept it with perfect probability (i.e., 1) and accordingly update its opinion unless $i$ is an originator.  
\vspace{-1mm}
\subsection{Epidemic Status of Agents} \label{subsec:epidemic_status}
\vspace{-1mm}
We model the evolution of false information propagation using a variant of the SIR model~\cite{Cho17}. The three states in the SIR model are defined based on the conditions associated with the expected belief or disbelief probabilities, $E_{b_i}$ and $E_{d_i}$, as follows:
\begin{itemize}
\item \emph{Susceptible} ($\mathcal{S}$): An agent is not sure of whether it believes false information or not. Agents in $\mathcal{S}$ have opinions with $E_b \leq 0.5$ and $E_d \leq 0.5$;
\item \emph{Infected} ($\mathcal{I}$): An agent believes false information as true with $E_b > 0.5$; and 
\item \emph{Recovered} ($\mathcal{R}$): An agent does not believe false information with $E_d > 0.5$. 
\end{itemize}
\vspace{-2mm}
\section{Numerical Results and Analysis} \label{sec:results}
\vspace{-1mm}
In this section, we describe metrics used for the experiments and detailed experimental setup. We also analyze the experimental results and discuss their overall trends.
\vspace{-1mm}
\subsection{Metrics} \label{subsec:metrics}
\begin{itemize}
\vspace{-1mm}
\item \emph{Agents' average opinion}: This metric shows the average value of an opinion with three dimensions, belief, disbelief, and uncertainty. Since the scenario is that belief indicates an supporting opinion for given false information, lower belief and higher disbelief are desirable while the degree of uncertainty will play a role in increasing/decreasing belief/disbelief.
\item \emph{Fraction of recovered agents} ($\mathcal{R}$): Based on the SIR model used in our work (see Section \ref{subsec:epidemic_status}), this metric refers to the fraction of agents which are in the status of the recovered ($\mathcal{R}(t)$) over all propagators. This metric indicates the fraction of agents that do not believe false information with $E_d > 0.5$.
\end{itemize}
\vspace{-2mm}
\subsection{Experimental Setup} \label{subsec:setup}
\begin{table}
\vspace{-3mm}
\caption{Key parameters and their default values}
\vspace{-2mm}
\centering
\begin{tabular}{|c|c|c|c|c|c|}
\hline
 param. & val. & param. & val. & param. & val.\\
\hline
$n$ & 1000 & $(tc_{\mu}, tc_{std})$ & $(0.5, 0.1)$ & $\gamma$ & 0.05\\
\hline
$s^*$ & $N \times 0.01$ & $(a_{\mu}, a_{std})$ & $(0.5, 0.1)$  & $n_r$ & 100 \\
\hline
\end{tabular}
\label{table_default_values}
\vspace{-3mm}
\end{table}
\begin{table}
\caption{Network dataset statistics}
\vspace{-2mm}
\centering
\begin{tabular}{|c|c|c|c|c|c|}
\hline
$N$ & 1033 & Ave. degree & 51.785 & Ave. path length & 2.949\\
\hline
$|\mathcal{E}|$ & 26747 & Modularity & 0.54 & Ave. clustering coeff. & 0.534 \\
\hline
\end{tabular}
\label{table_network statistics}
\vspace{-5mm}
\end{table}
For the network topology, we use an ego-Facebook dataset~\cite{SNAP} which gives a fully connected undirected network described by Table \ref{table_network statistics}. To disseminate the original public information by a given set of originators, we seed 1 \% of the total nodes, denoted by $s^*$, which is 11 nodes in the given network and model the rest of the nodes (i.e., 1022) as propagators. We consider opinion decay factor $\gamma$ set to 0.05 and mean prior belief $a_{\mu} = 0.5$ with the standard deviation $a_{std}=0.1$, implying that both prior belief and prior disbelief are fairly same. Agents' topic competence is assigned with the mean $tc_{\mu}=0.5$ with the standard deviation $tc_{std}=0.1$. The key design parameters and their default values used for our experiments are summarized in Table \ref{table_default_values}. The data points shown in the results are the average values based on the collected data from simulation runs $n_r=100$.

Information propagation proceeds as follows. A given number of originators, $s^*$, is initiated with the set of opinions, $\hat{Q}$ as described in Section \ref{subsec:information_model}. After the originators form their opinions by processing the four types of evidence, PV, PC, CV, and CN, they propagate their opinions to their neighbors. Accordingly, the neighbors also forward their updated opinions to their neighbors based on Eq. \eqref{consensus_op}. Note that each agent has a chance to disseminate its opinion to its neighbors. We discuss the effect of the key design parameter values on the given metrics in Section \ref{subsec:metrics} in the following section.

\begin{figure*}[!ht]
\vspace{-4mm}
\centering
\subfloat[$b$]{\includegraphics[width=1.7in, height=1in]{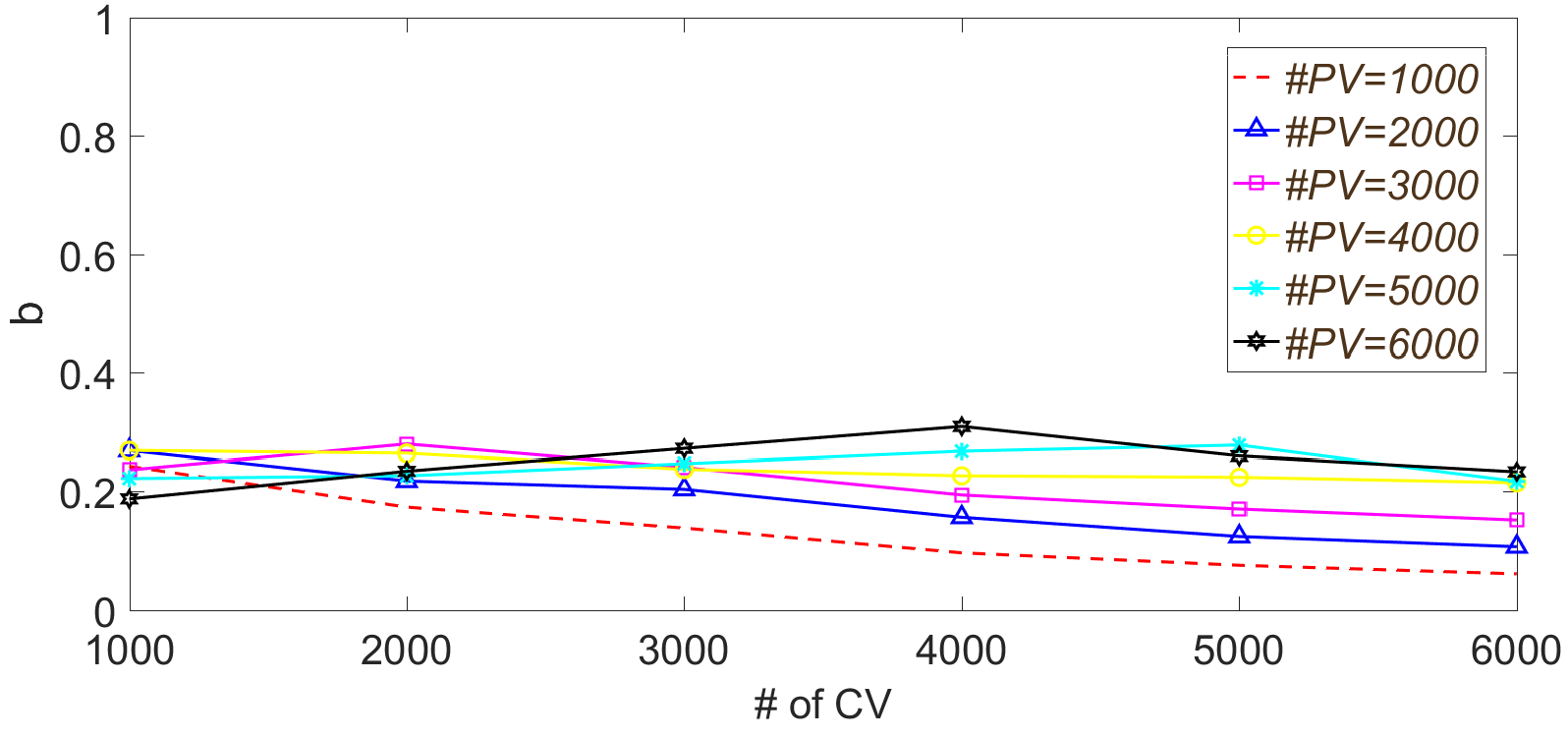}\label{fig_b_pcv}}
\hfil
\subfloat[$d$]{\includegraphics[width=1.7in, height=1in]{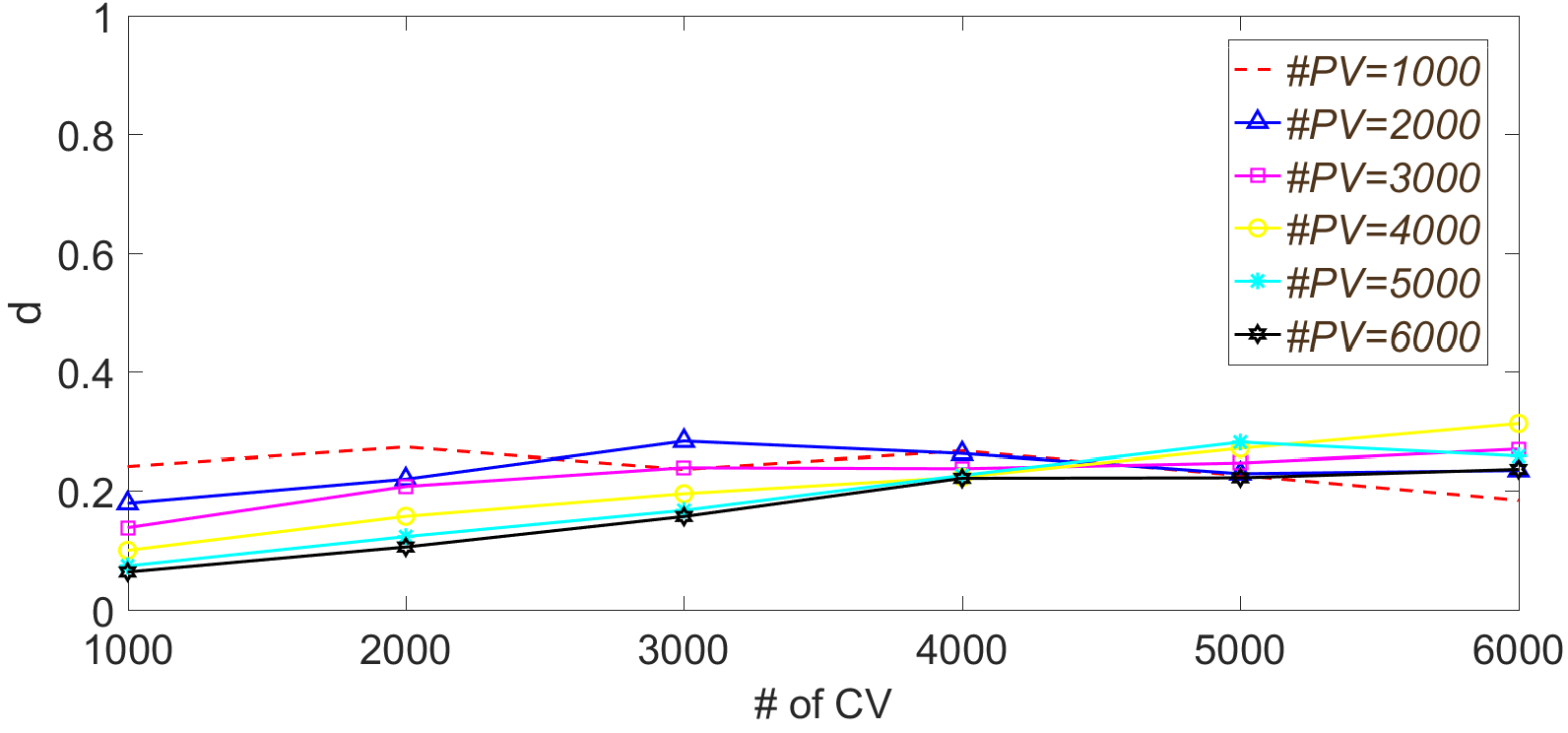}\label{fig_d_pcv}}
\hfil
\subfloat[$u$]{\includegraphics[width=1.7in, height=1in]{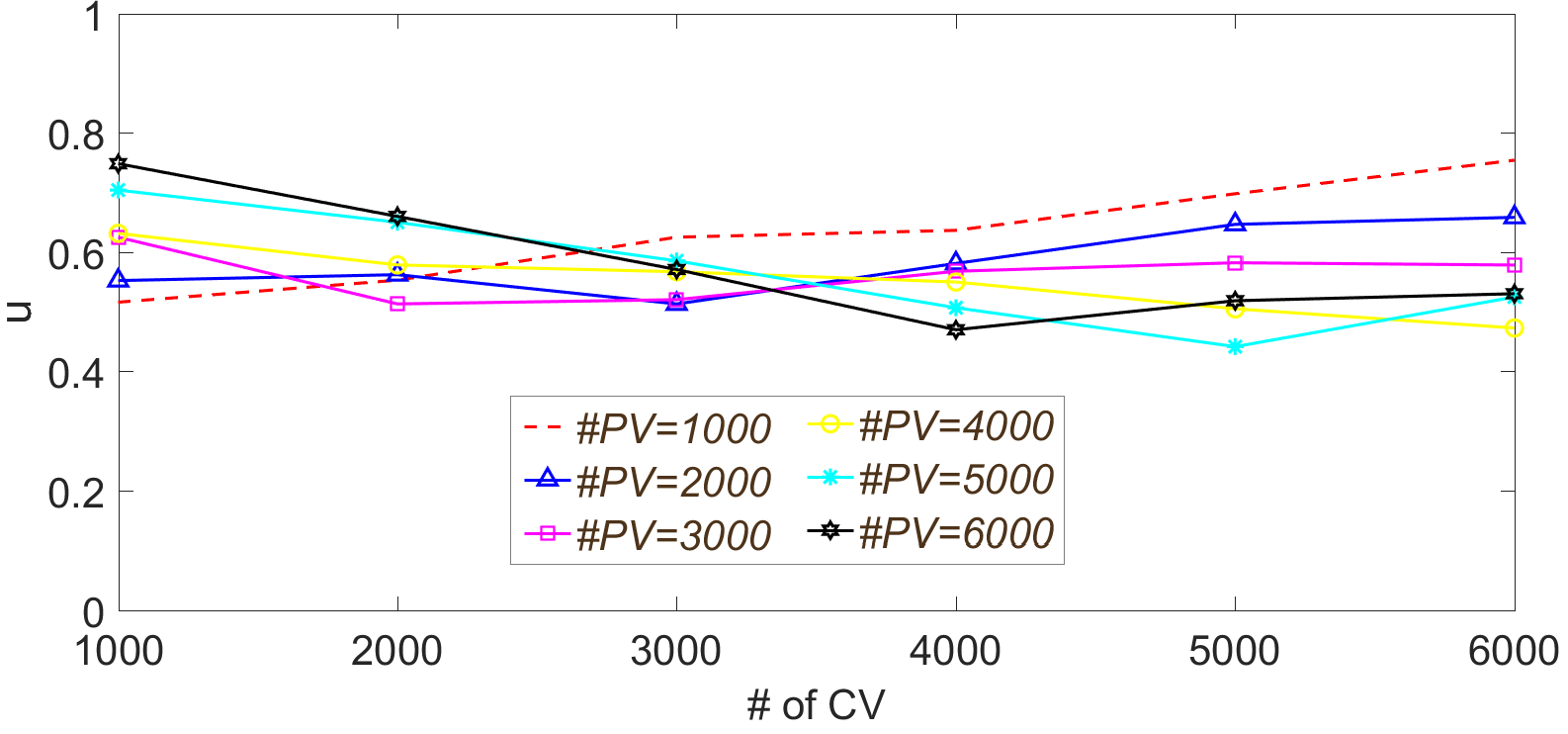}\label{fig_u_pcv}}
\hfil
\subfloat[$\mathcal{R}$]{\includegraphics[width=1.7in, height=1in]{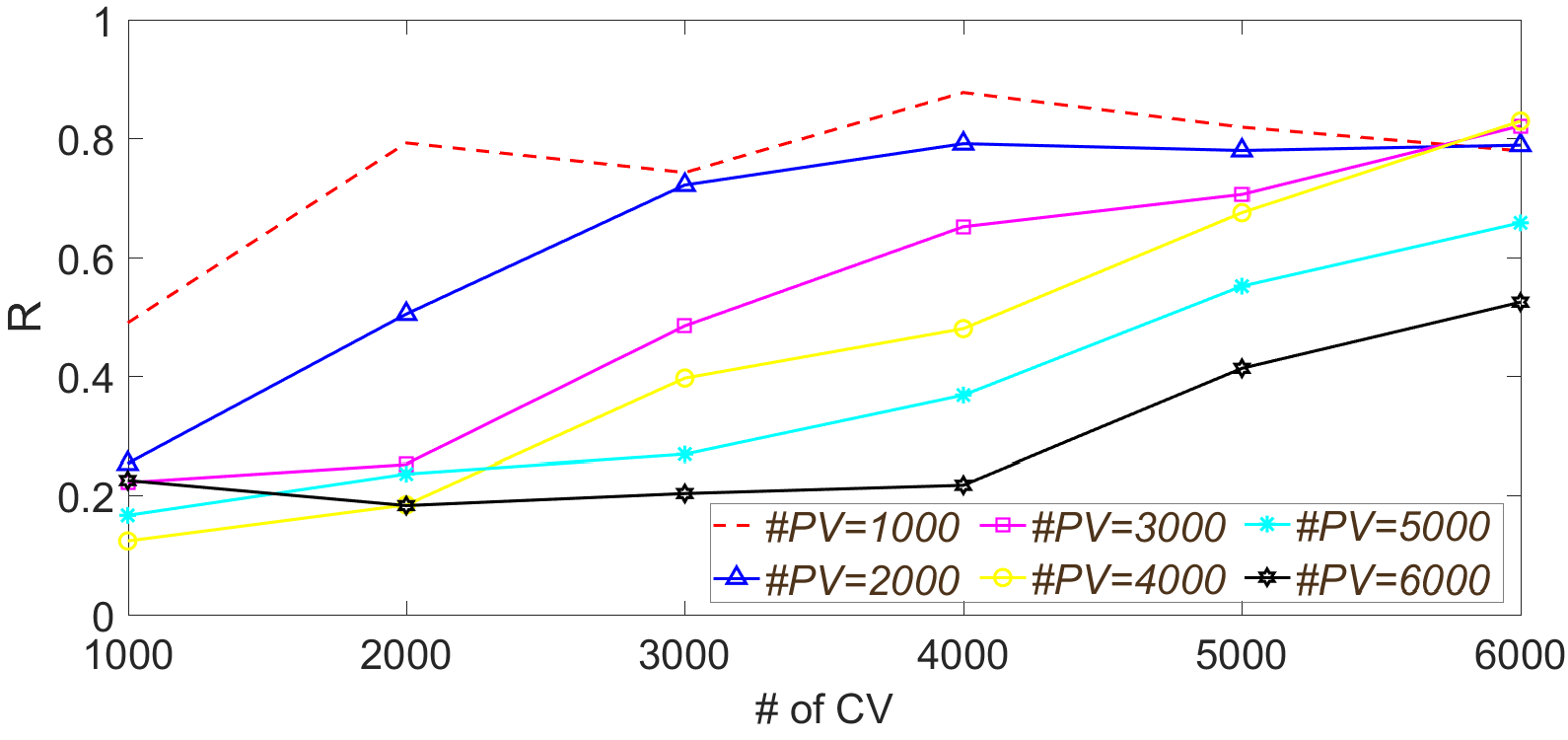}\label{fig_r_pcv}}
\caption{Effect of varying the amount of valuable information}
\label{fig_PCV}
\vspace{-4mm}
\end{figure*}

\begin{figure*}[!ht]
\vspace{-4mm}
\centering
\subfloat[$b$]{\includegraphics[width=1.7in, height=1in]{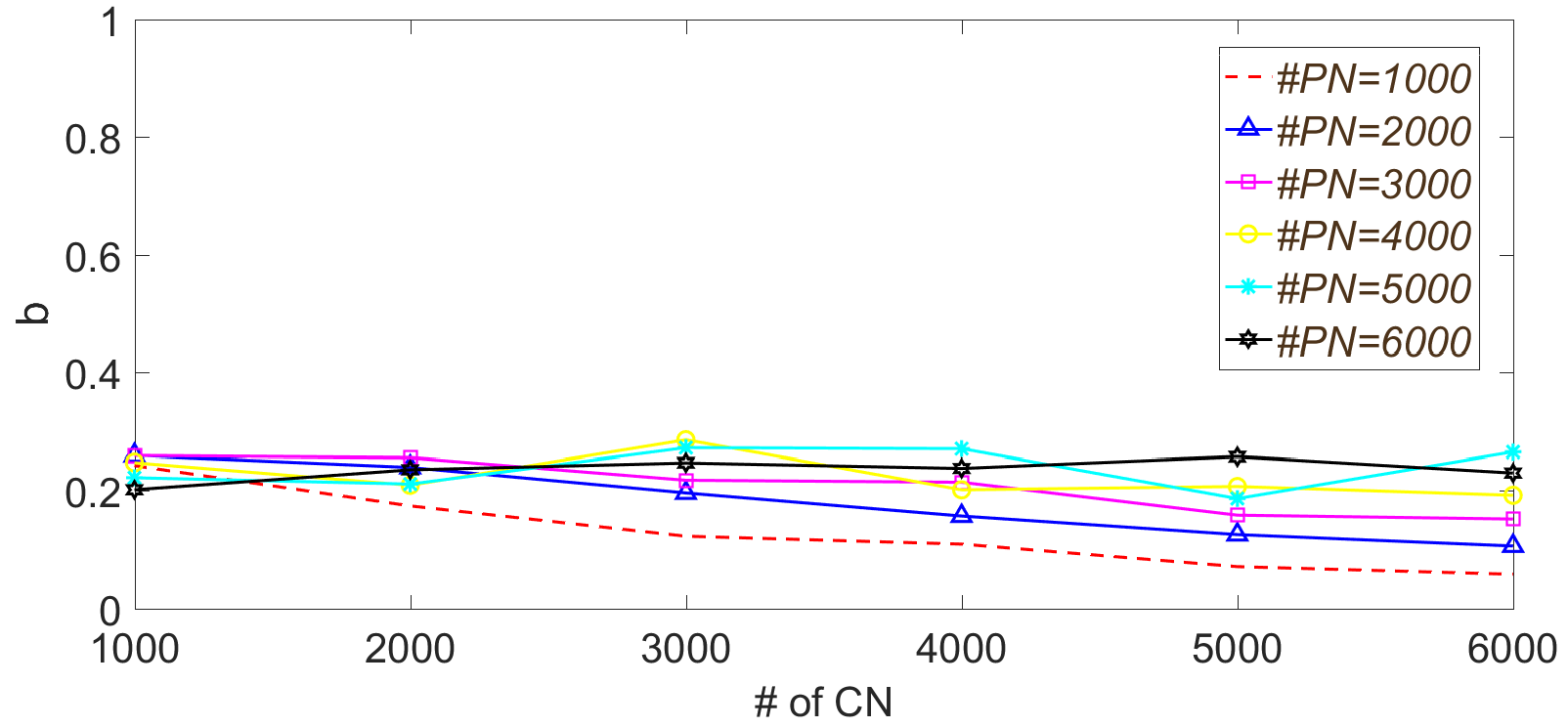}\label{fig_b_pcn}}
\hfil
\subfloat[$d$]{\includegraphics[width=1.7in, height=1in]{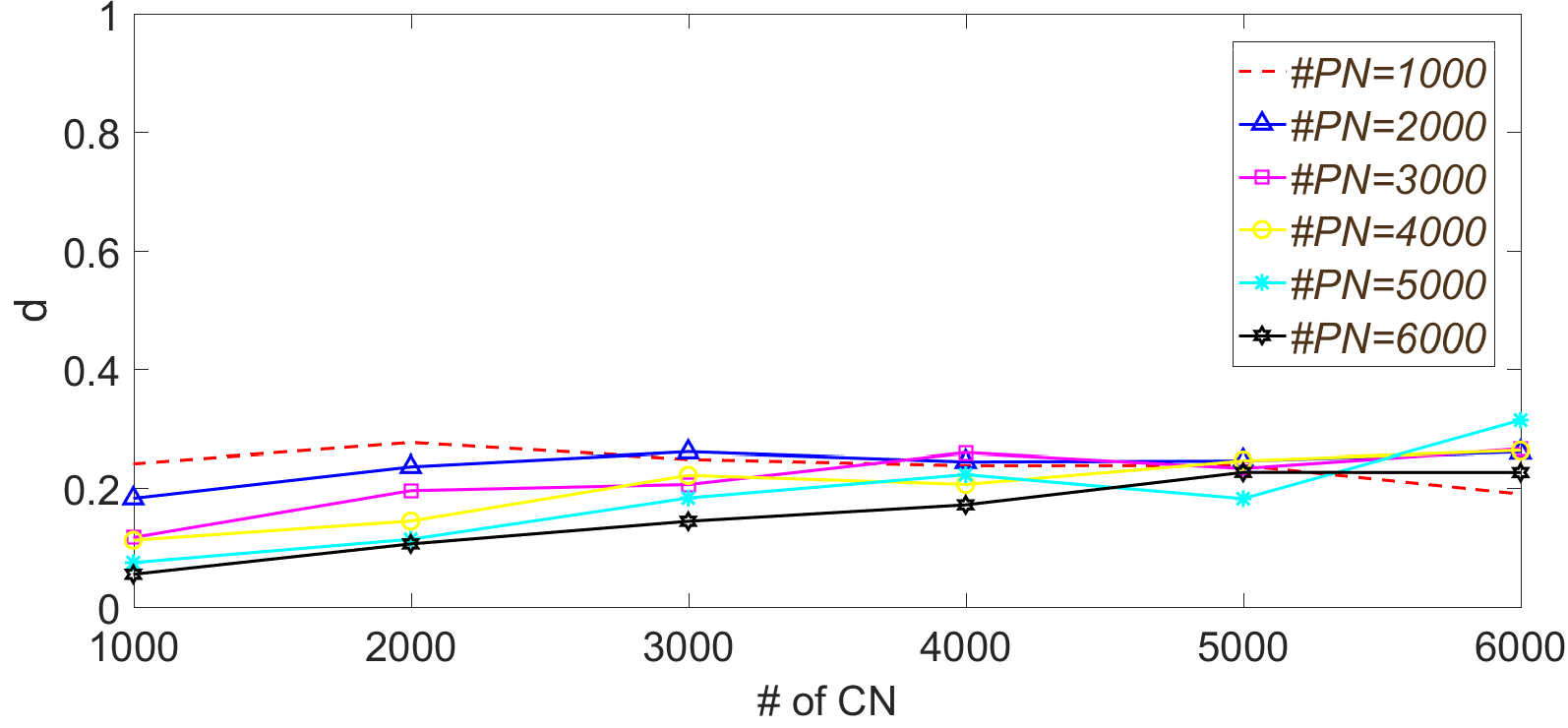}\label{fig_d_pcn}}
\hfil
\subfloat[$u$]{\includegraphics[width=1.7in, height=1in]{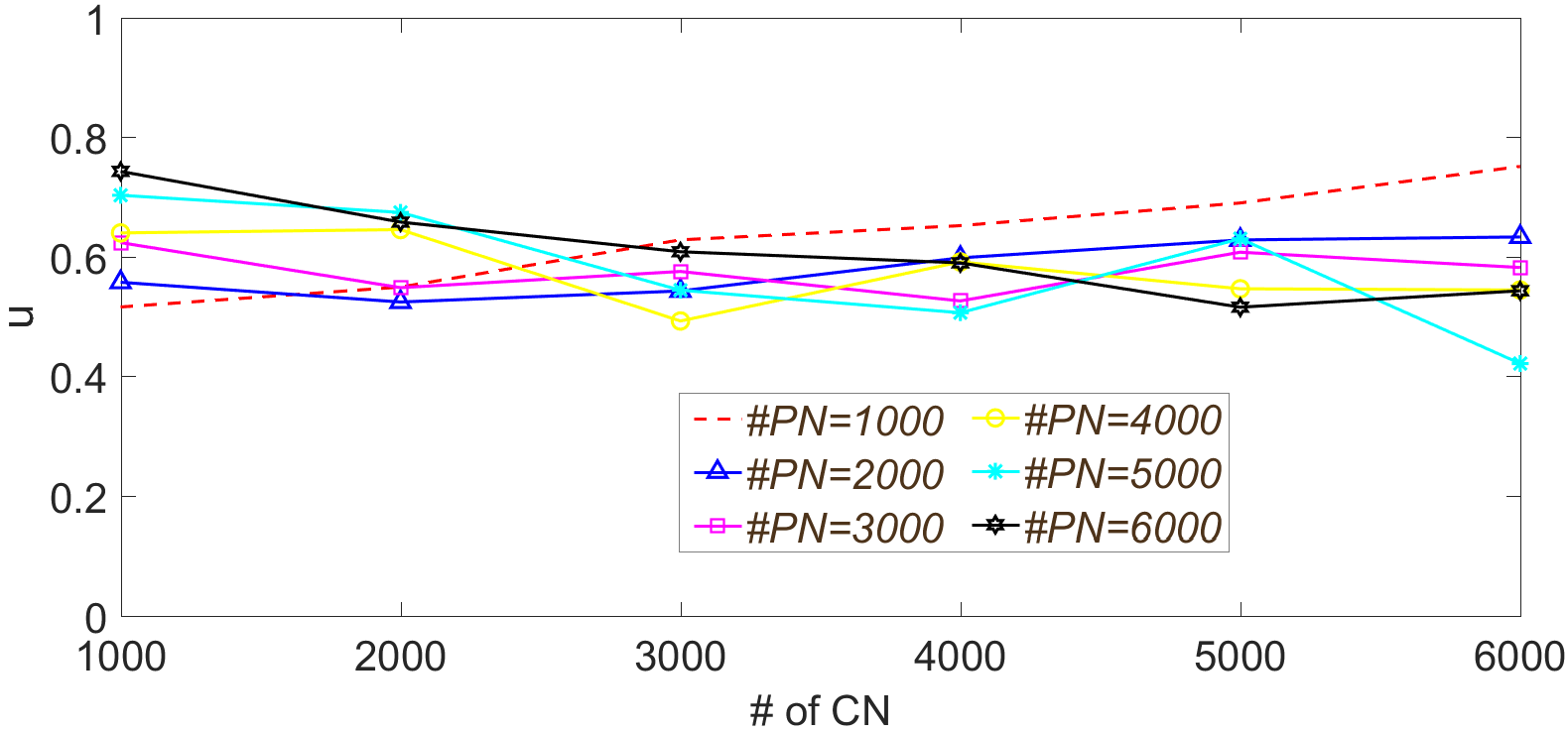}\label{fig_u_pcn}}
\hfil
\subfloat[$\mathcal{R}$]{\includegraphics[width=1.7in, height=1in]{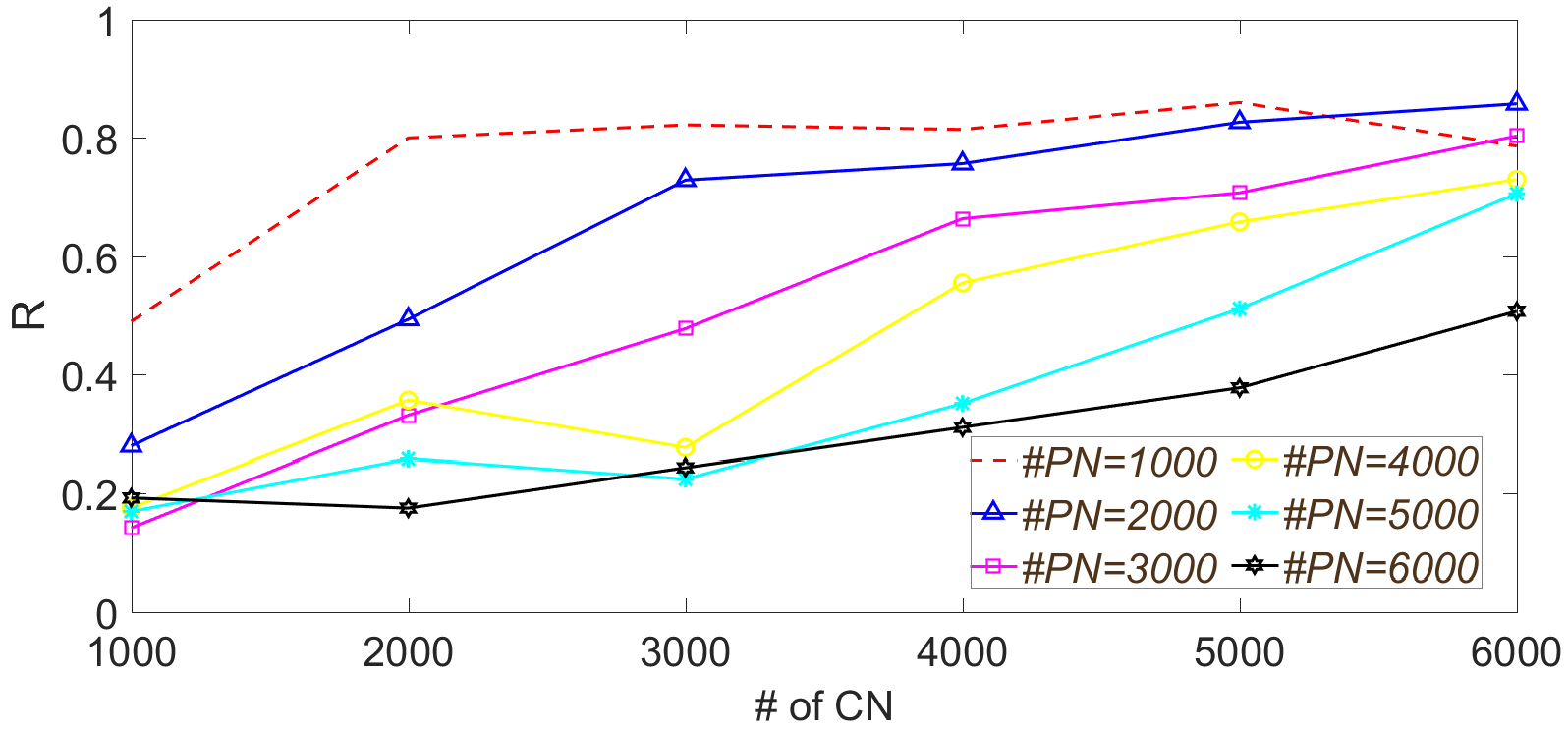}\label{fig_r_pcn}}
\caption{Effect of varying the amount of noisy information}
\label{fig_PCN}
\vspace{-4mm}
\end{figure*}

\begin{figure*}[!ht]
\vspace{-4mm}
\centering
\subfloat[$b$]{\includegraphics[width=1.7in, height=1in]{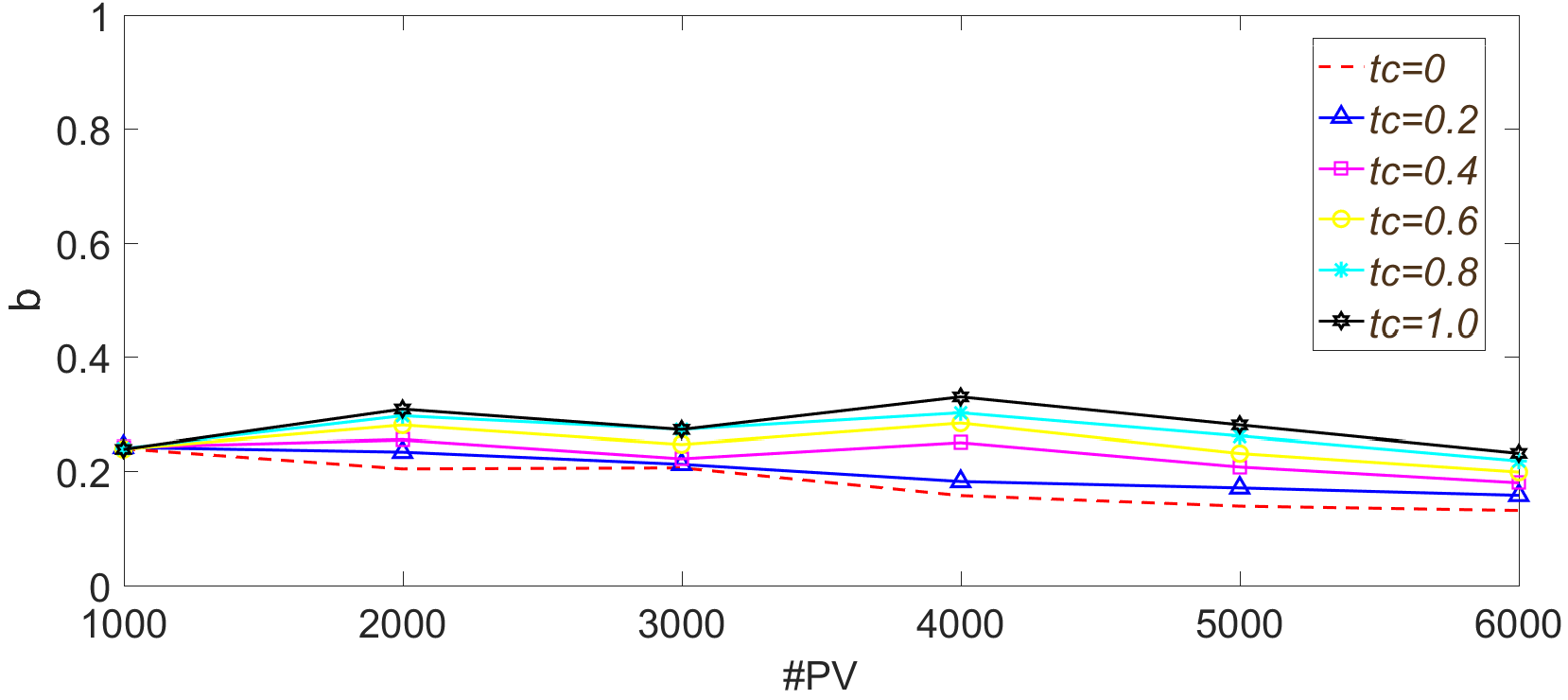}\label{fig_b_tc_pv}}
\hfil
\subfloat[$d$]{\includegraphics[width=1.7in, height=1in]{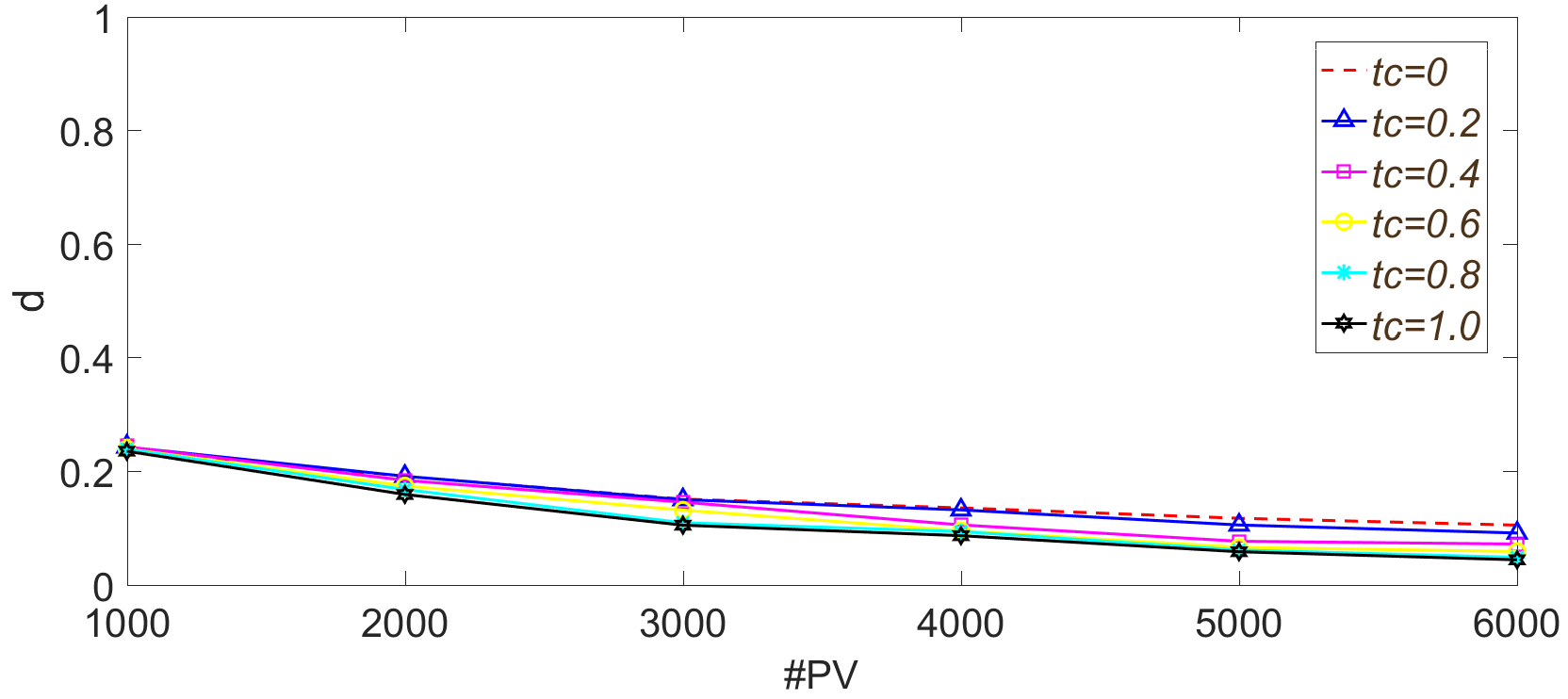}\label{fig_d_tc_pv}}
\hfil
\subfloat[$u$]{\includegraphics[width=1.7in, height=1in]{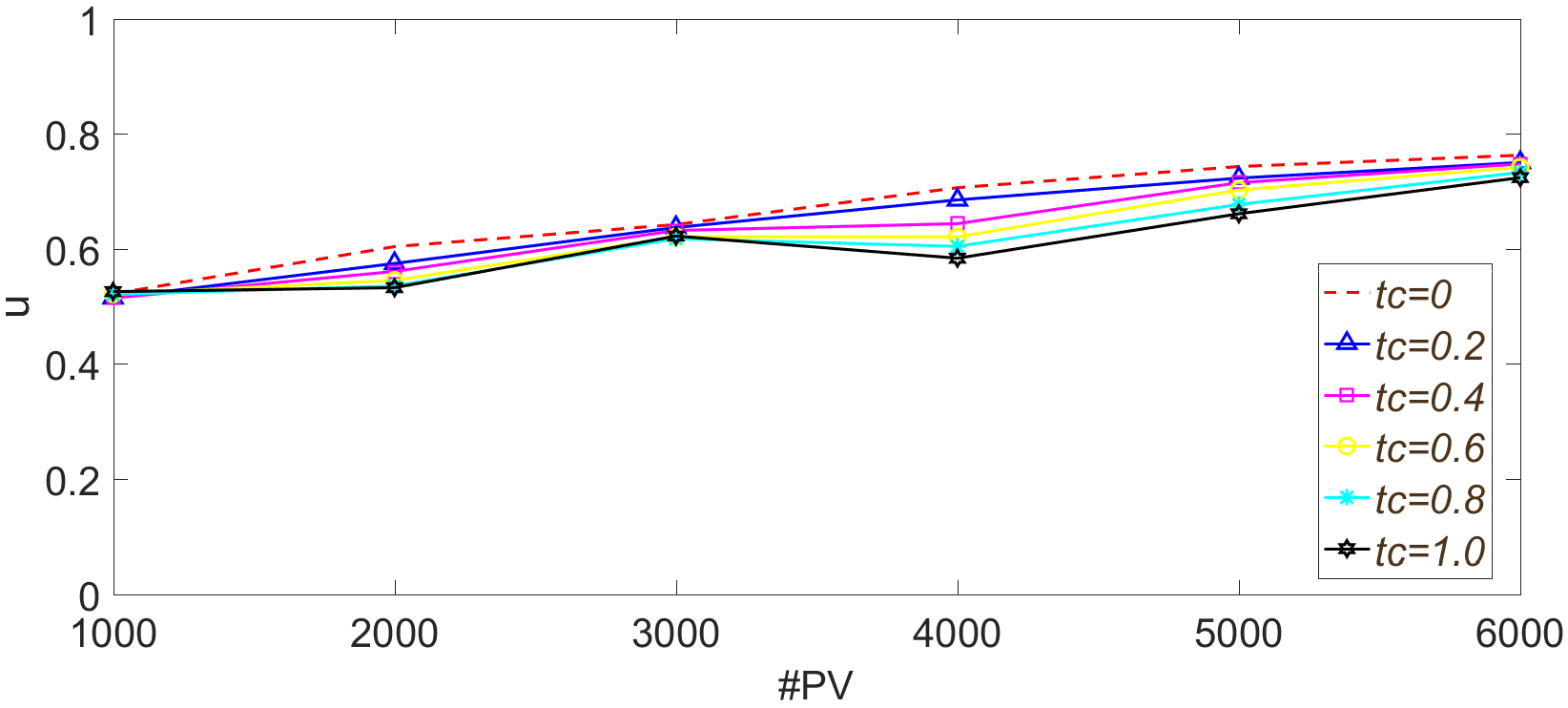}\label{fig_u_tc_pv}}
\hfil
\subfloat[$\mathcal{R}$]{\includegraphics[width=1.7in, height=1in]{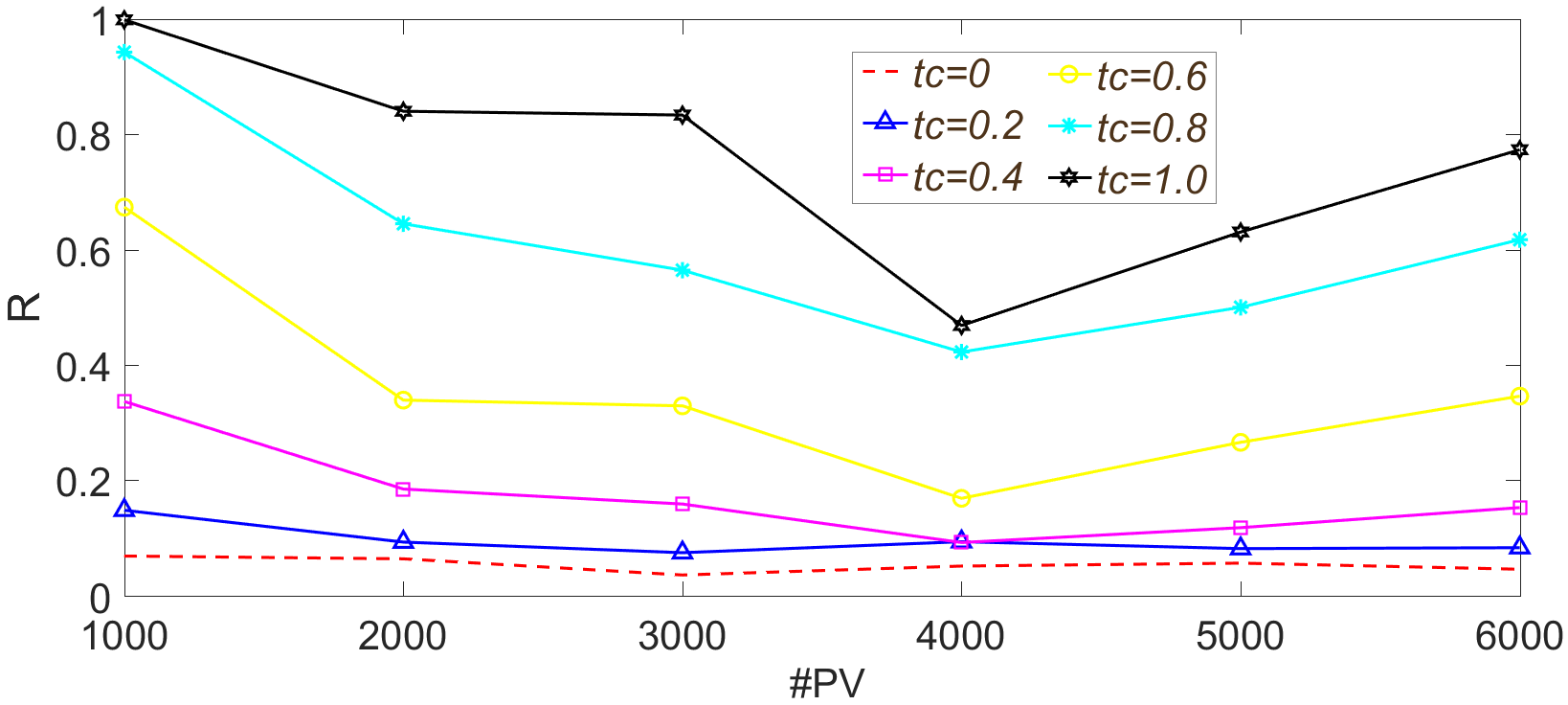}\label{fig_r_tc_pv}}
\caption{Effect of the topic competence under varying the amount of valuable $pro$ evidence (PV)}
\label{fig_TC_PV}
\vspace{-4mm}
\end{figure*}

\begin{figure*}[!ht]
\vspace{-4mm}
\centering
\subfloat[$b$]{\includegraphics[width=1.7in, height=1in]{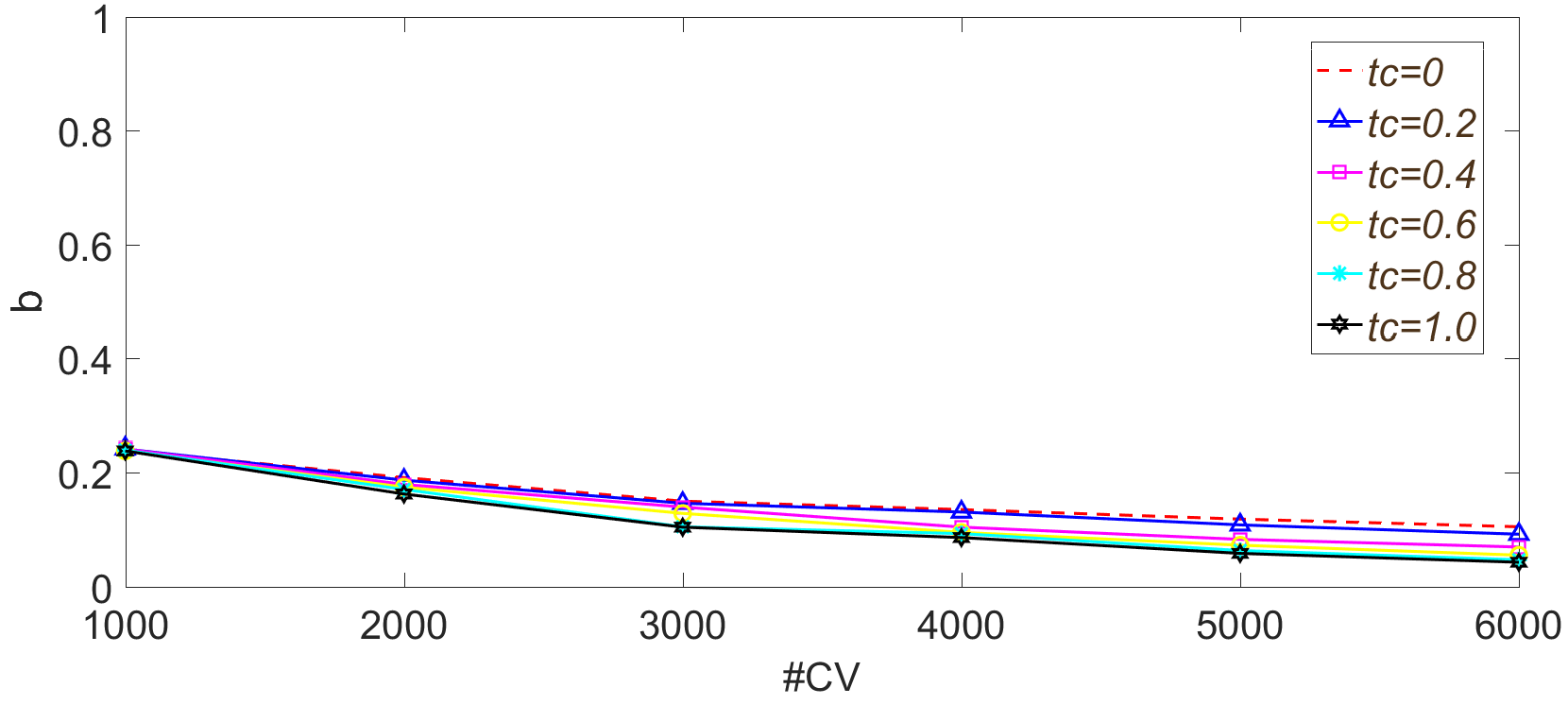}\label{fig_b_tc_cv}}
\hfil
\subfloat[$d$]{\includegraphics[width=1.7in, height=1in]{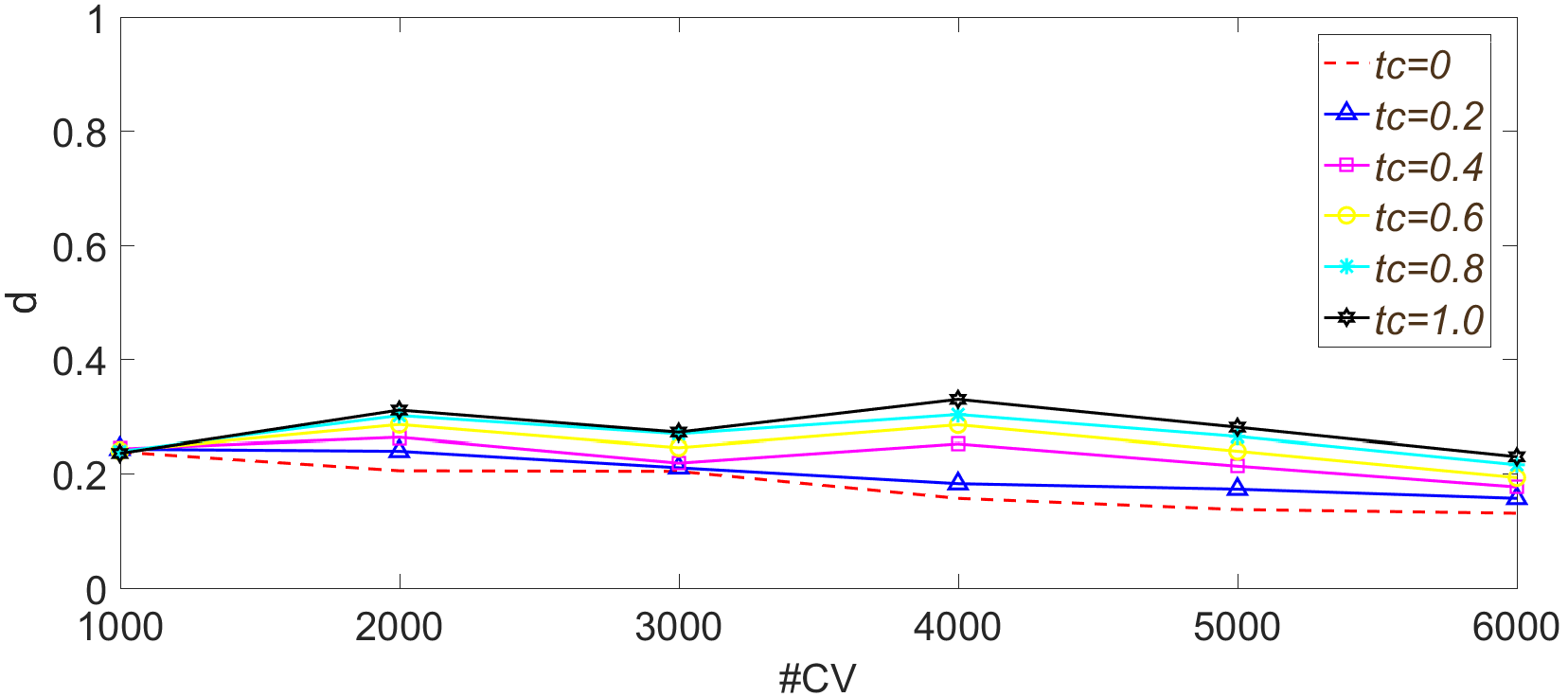}\label{fig_d_tc_cv}}
\hfil
\subfloat[$u$]{\includegraphics[width=1.7in, height=1in]{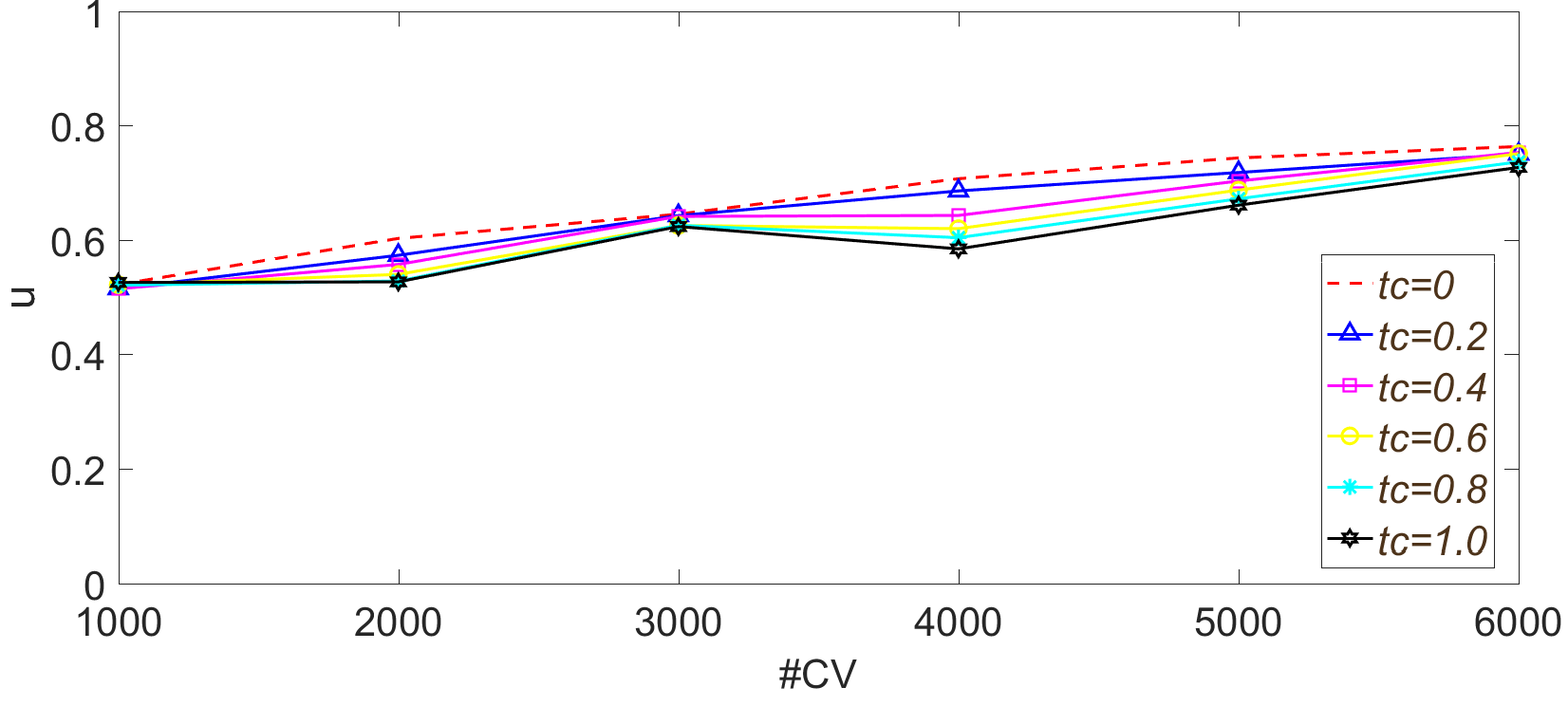}\label{fig_u_tc_cv}}
\hfil
\subfloat[$\mathcal{R}$]{\includegraphics[width=1.7in, height=1in]{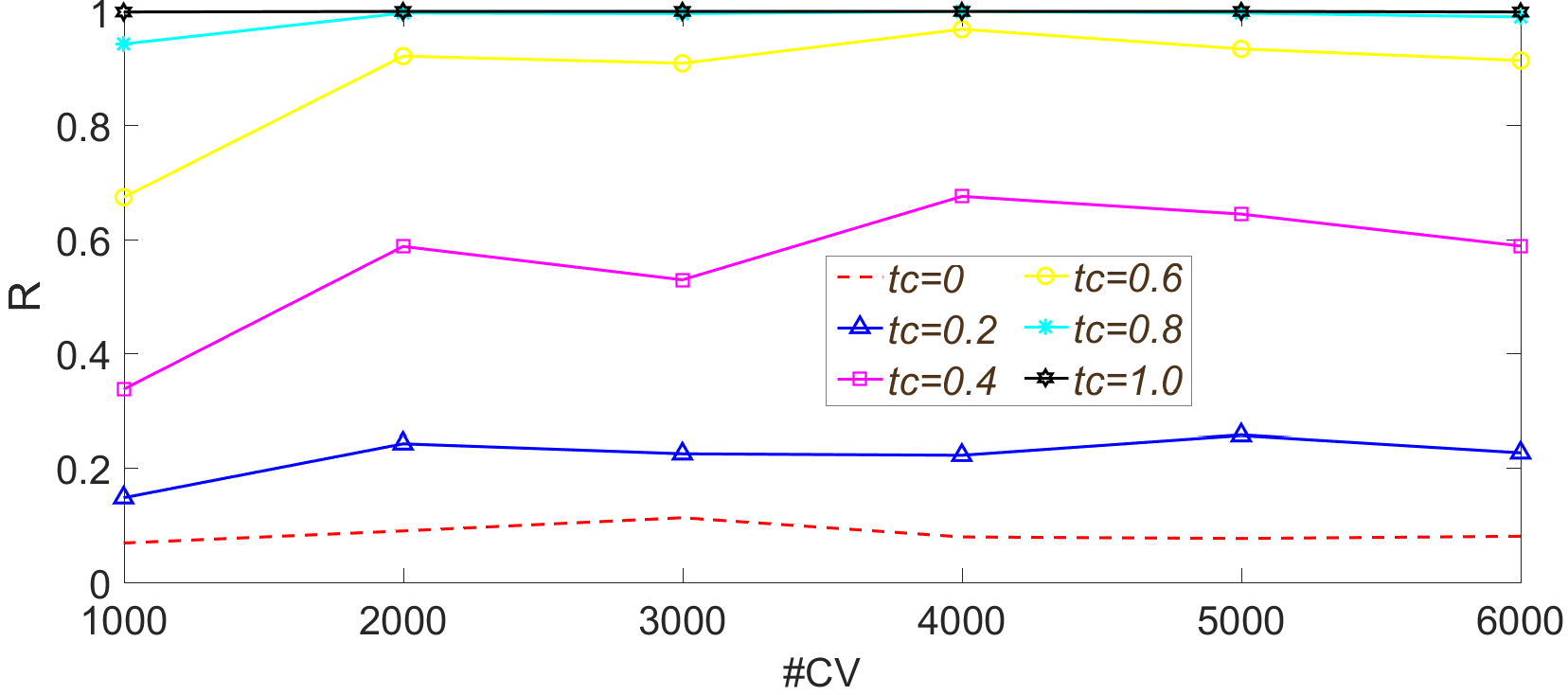}\label{fig_r_tc_cv}}
\caption{Effect of the topic competence under the amount of valuable $con$ evidence (CV)}
\label{fig_TC_CV}
\vspace{-4mm}
\end{figure*}
\vspace{-1mm}
\subsection{Results and Analysis} \label{subsec:results_analysis}
In this section, we conduct sensitivity analysis by varying the values of key design parameters, mainly including varying the ratio of information types (i.e., \# of PV, PN, CV, and CN) and the degree of agents' topic competence.
\subsubsection{\bf Effect of Varying the Amount of Valuable Information (PV and CV)} Fig. \ref{fig_PCV} shows how the amount of valuable evidence (i.e., \#PV and \#CV) affects agents' mean opinion in terms of three dimensions, denoted by $b$, $d$, and $u$, and the fraction of recovered agents ($\mathcal{R}$). In Fig. \ref{fig_PCV} (a) showing the mean $b$, the increase of \#CV decreases $b$ because more uncertain evidence is introduced by agents' imperfect topic competence (i.e., $tc_{\mu}=0.5$). However, as \#PV increases, $b$ increases with relatively less uncertain evidence, as seen in Fig. \ref{fig_PCV} (c). In addition, the increasing trends of $b$ is observed as \#CV increases with higher \#PV ($\geq 5000$)  because of the decreasing uncertain evidence $u$ based on the increased amount of valuable evidence. However, with higher \#CV ($\geq 4000$), $b$ decreases again due to the increased uncertain evidence introduced by imperfect competence used to detect noisy or valuable evidence. In Fig. \ref{fig_PCV} (b), we show the effect of varying \#CV and \#PV in $d$. As expected, higher \#PV shows the lowest $d$ while the increase of \#CV increases $d$ overall. In Fig. \ref{fig_PCV} (b), lower \#PV and/or higher \#CV increases $d$ which is intuitively true. In Fig. \ref{fig_PCV} (c), we show $u$ as \#CV and \#PV increase. Under relatively low \#PV (i.e., \#PV $\leq 2000$), we can see increasing uncertainty with higher \#CV because of the imperfect topic competence. However, under relatively higher \#PV, higher \#CV decreases $u$ due to the increased effect of $d$. In particular, when \#PV$=6000$, the minimum point of $u$ at \#CV$=4000$, which is generated based on the tradeoff between $b$ and $d$ because of $b+d+u=1$. Finally, Fig. \ref{fig_PCV} (d) shows the fraction of recovered agents, $\mathcal{R}$, where an agent is identified as {\em recovered}, meaning that the agent does not believe false information, $b$, with $E_d > 0.5$. From this figure, we can clearly observe the positive effect of \#CV, showing that higher \#CV increases $\mathcal{R}$, because its misdetection can increase uncertainty which is also considered in calculating expected disbelief in Eq. \eqref{eq:expectation}. In addition, higher \#PV is also clear by decreasing $\mathcal{R}$. The noticeable finding is that the imperfect topic competence increases uncertain evidence which can allow agents to make right decisions (disbelieving false information) because they are not at least biased for false information $b$. However, the increased uncertainty does not allow $d$ or $\mathcal{R}$ to increase further. 
\subsubsection{\bf Effect of Varying the Amount of Noisy Information (PN and CN)} Fig. \ref{fig_PCN} shows how the noisy information (i.e., \#CV and \#CN) affects $b$, $d$, $u$, and $\mathcal{R}$ when agents are given with the mean topic competence, $tc=0.5$. Very similar to the trends observed in Fig. \ref{fig_PCV}, the effect of \#CN is clearer in $b$ with lower \#PN and in $d$ with higher \#PN. That is, higher \#PN significantly mitigates the positive effect of higher \#CN, detecting half valuable con and half noisy con which lead to higher $d$. Overall the effect of noisy information is very similar to that of valuable information although the effect of valuable information slightly shows the better performance in $\mathcal{R}$. Now we discuss the effect of varying agents' topic competence as below.
\subsubsection{\bf Effect of Varying the Amount of Agents' Topic Competence under Varying the Amount of PV} Fig. \ref{fig_TC_PV} shows how agents process their opinions and update them based on their different topic competence with respect to varying \#PV when \#CV, \#CN, and \#PN are fixed at 1000. From this figure, we find very interesting results because the trends of $b$ and $d$ look counter-intuitive. But the reasons can be well explained in terms of the amount of uncertain evidence diagnosed based on agents' topic competence given. To be specific, in $b$ shown in Fig. \ref{fig_TC_PV} (a), the lowest topic competence (i.e., $tc=0$) shows the lowest $b$ while the highest topic competence (i.e., $tc=1$) shows the highest $b$ because $tc=0$ takes all $b$ and $d$ evidence (i.e., \#PV and \#CV) as uncertain evidence, $u$, resulting in too high uncertainty generated, as shown in $u$ shown in \ref{fig_TC_PV} (c). Similarly, the effect of significantly increased $u$ is observed in $b$ (\ref{fig_TC_PV}  (a)), showing the highest $b$ with $tc=1$. The increased $b$ with $tc=1$ also reduces $d$ as $b+d+u=1$. When $tc=1$, only noisy information (1000 \#PN and 1000 \#CN) is counted as uncertain evidence while others will be either $b$ or $d$. Thus, uncertainty with $tc=1$ is the lowest. That is, compared to the increased $tc$ that reduces agents' prior belief (see Eq. \eqref{eq:base}) and lessens $b$ but increases $d$, the largeness of increased uncertain evidence, $u$, exceeds the effect of reduced prior belief, resulting in higher $b$ and lower $d$ with high $tc$ where \#PV is larger than any other evidence types.
\subsubsection{\bf Effect of Varying the Amount of Agents' Topic Competence under Varying the Amount of CV} Lastly, Fig. \ref{fig_TC_CV} shows the effect of agents' topic competence with respect to varying \#CV when \#CN, \#PV, and \#PN are set to 1000. Recall that based on Algorithm \ref{algo: mapping_evidence_to_opinion}, the correct detection of \#CV means that the evidence is detected as valuable con information (i.e., true information) while misdetection of \#CV means that the evidence is detected as uncertain evidence. Unlike Fig. \ref{fig_TC_PV} showing that higher $tc$ generates more $b$ and less $d$ as \#PV increases, the effect of \#CV follows the intuitive results such that higher $tc$ generates more $d$ and less $b$, leading to higher $\mathcal{R}$. As mentioned earlier, this trend is reasonable because higher \#CV means more valuable evidence to support $con$ which corresponds to $d$ while generated uncertain evidence due to imperfect topic competence from given \#CV can increase $\mathcal{R}$ as well.
\vspace{-5mm}
\section{Conclusion} \label{sec:conclusion}
\vspace{-1mm}
The key findings are summarized as follows:
\begin{itemize}
\item An appropriate amount of valuable information (e.g., \#CV) exists to maximize valuable evidence to support truth while minimizing uncertain evidence.
\item The amount of uncertain evidence can help when agents are not at least biased favoring for false information (i.e., $b$), meaning that agents' base rate, $a_i$, does not exceed 0.5, resulting in higher $\mathcal{R}$.
\item When the amount of false evidence (e.g., \#PV) exceeds that of true evidence (e.g., \#CV), higher $tc$ may increase the amount of false information, $b$. However, since higher $tc$ can reduce the effect of false information, $b$, by using lower prior belief in the false information, higher $tc$ can achieve higher $\mathcal{R}$ even with more valuable evidence supporting false information with the help of uncertain evidence which is interpreted as not supporting the false information.
\item When the amount of true evidence (e.g., \#CV) exceeds that of false evidence (e.g., \#PV), higher $tc$ helps increase $d$ and accordingly $\mathcal{R}$. This is because even misdetection of true evidence can increase the amount of uncertain evidence which can be well utilized to increase expected disbelief, leading to higher $\mathcal{R}$ with the reduced bias towards false information with higher $tc$.    
\end{itemize} 
The future work directions of this research can follow: (i) the investigation on the effect of centrality types of information originators in various network topologies; (ii) the improvement of designing agents' prior belief (i.e., base rate) to dynamic prior belief based on all prior evidence; and (iii) the development of uncertainty detector by examining text mining techniques which can be used for the validation of the presented opinion and information model.
 
\section*{Acknowledgment}
This work was partially supported by the U.S. Army Research Laboratory under Cooperative Agreement No. W911NF-09-2-0053; The views and conclusions contained in this document are those of the authors and should not be interpreted as representing the official policies, either expressed or implied, of ARL, NSF, or the U.S. Government. The U.S. Government is authorized to reproduce and distribute reprints for Government purposes notwithstanding any copyright notation here on.
\vspace{-1mm}
\bibliographystyle{IEEETranSN}
{\small
\bibliography{icc-references}
}

\end{document}